\documentclass[a4paper]{article}

\usepackage{icrc2013}

\title{A Compact High Energy Camera for the Cherenkov Telescope Array}

\shorttitle{CHEC}

\authors{
M.~K.~Daniel$^{1}$,
R.~J.~White$^{2}$,
D.~Berge$^{3}$,
J.~Buckley$^{4}$,
P.~M.~Chadwick$^{5}$, 
G.~Cotter$^{6}$,
S.~Funk$^{7}$,
T.~Greenshaw$^{1}$,
N.~Hidaka$^{8}$,
J.~Hinton$^{2}$,
J.~Lapington$^{2}$,
S.~Markoff$^{3}$,
P.~Moore$^{4}$,
S.~Nolan$^{5}$,
S.~Ohm$^{2}$,
A.~Okumura$^{8,2}$,
D.~Ross$^{2}$,
L.~Sapozhnikov$^{7}$,
J.~Schmoll$^{5}$,
P.~Sutcliffe$^{1}$,
J.~Sykes$^{2}$,
H.~Tajima$^{8}$,
G.~S.~Varner$^{9}$,
J.~Vandenbroucke$^{7}$,
J.~Vink$^{3}$,
D.~Williams$^{10}$
for the CTA Consortium.
}

\afiliations{
$^1$ University of Liverpool, UK.\\
$^2$ University of Leicester, UK.\\
$^3$ University of Amsterdam, Netherlands.\\
$^4$ Washington University at St Louis, USA. \\
$^5$ University of Durham, UK.\\
$^6$ University of Oxford, UK.\\
$^7$ Stanford University and SLAC National Accelerator Laboratory, USA.\\
$^8$ Solar-Terrestrial Environment Laboratory, Nagoya University, Japan.\\
$^9$ University of Hawaii, USA. \\
$^{10}$ University of California, Santa Cruz. USA.
}

\email{Michael.Daniel@liverpool.ac.uk}

\abstract{
The Compact High Energy Camera (CHEC) is a camera-development project involving 
UK, US, Japanese and Dutch institutes for the dual-mirror Small-Sized 
Telescopes (SST-2M) of the Cherenkov Telescope Array (CTA). Two CHEC prototypes, 
based on different photosensors are funded and will be assembled and tested in 
the UK over the next $\approx$18 months. CHEC is designed to record flashes of 
Cherenkov light lasting from a few to a hundred nanoseconds, with typical RMS 
image width and length of
$\sim0.2^{\circ}\times1.0^{\circ}$, and has a 9$^{\circ}$ field of view. The
physical camera geometry is dictated by the telescope optics: a curved focal
surface with radius of curvature 1~m and diameter $\sim$35\,cm is required. 
CHEC is designed to work with both the ASTRI and GATE SST-2M telescope 
structures and will include an internal LED flasher system for calibration. The 
first CHEC prototype will be based on multi-anode photomultipliers (MAPMs) and 
the second on silicon photomultipliers (SiPMs or MPPCs). The first prototype 
will soon be installed on the ASTRI SST-2M prototype structure on Mt. Etna.
}


\begin{document}
\maketitle

\section{Introduction}\label{sec:intro}
The highest energy photons are large, bright, but rare events so for CTA to 
achieve sensitivity a sparse array of Small Size Telescopes (SSTs) with large
fields of view is required. 
A dual-mirror SST solution allows less expensive small plate scale focal plane 
detectors to be used, the resulting camera savings means a greater number of 
SST-2Ms can be built allowing a greater coverage on the ground. 
The Compact High Energy Camera (CHEC) is a camera-development project 
specifically with both the ASTRI \cite{bib:DiPierro} and GATE \cite{bib:Zech} 
prototype SST-2M structures.  
Two CHEC prototypes, one based on multi-anode photomultipliers
(MAPMs) the other on silicon photomultipliers (SiPMs), are 
funded to allow competing technologies to be compared. The results of this
prototyping phase will then be merged into the production telescopes for CTA.


\section{CHEC}\label{sec:CHEC}

 \begin{figure*}[!t]
  \centering
  \includegraphics[trim=0cm 3.5cm 0cm 1.5cm, clip=true, width=\textwidth]{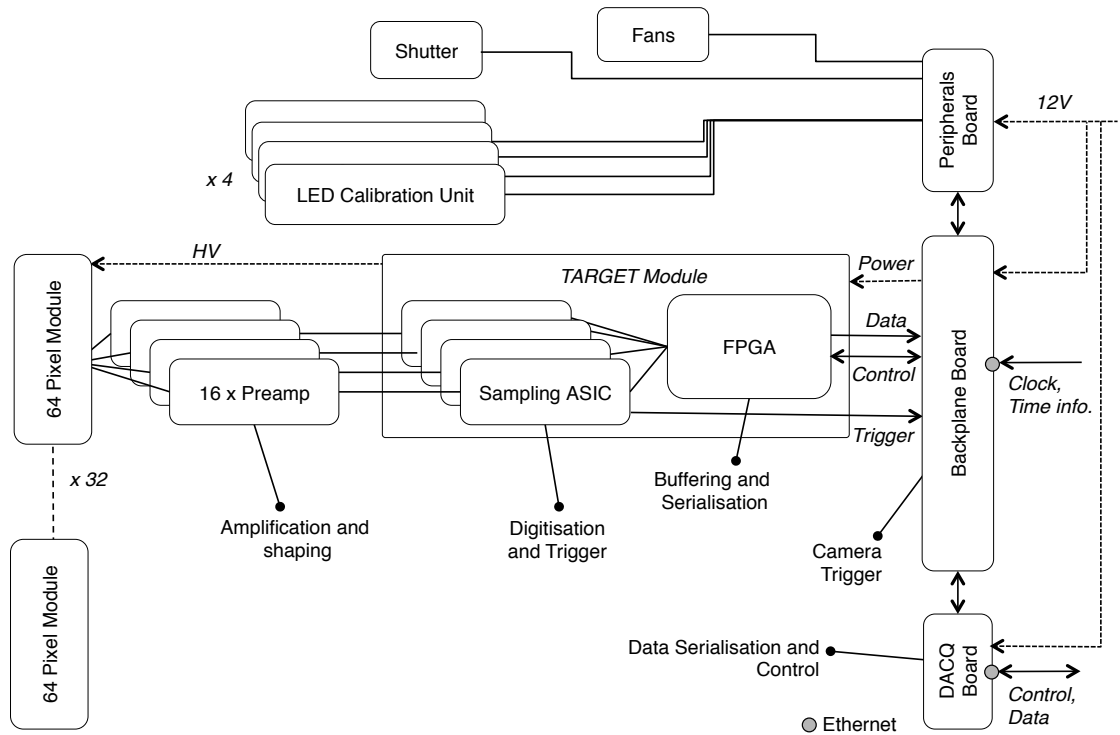}
  \caption{
           The CHEC concept
          }
  \label{fig:concept}
 \end{figure*}

The design concept for CHEC is given in Figure~\ref{fig:concept}.
\begin{itemize}
\item {\bf Photosensors - } MAPMs and SiPMs will be used for CHEC-M and
  CHEC-S respectively.
\item {\bf Preamplifiers - } Signal amplification and shaping to
  optimise camera triggering and readout. 
\item {\bf TARGET Modules - } 64-channel signal capture modules based
  on the TARGET ASIC for data digitisation, read out, pixel-level
  triggering and slow control.
\item {\bf Backplane Board - } All 32 TARGET Modules plug directly into
  a large backplane PCB that provides: camera-level triggering, clock
  distribution, communication with the TARGET Modules and routing to
  the DACQ board.
\item {\bf DACQ Board - } High-speed serial data from the TARGET
  Modules are routed to the outside world. Also provides control for camera 
  `peripherals' such as the lid and calibration systems.
\item {\bf Peripherals Board - } An interface to the various slow control 
      peripherals, eg calibration units, lid/shutter motor controller, ambient 
      light sensors, etc.
\item {\bf Mechanics - } The internal support structure, cooling
  system, enclosure and lid/shutter.
\end{itemize}
The physical camera geometry is dictated by the telescope optics: a curved 
focal surface with radius of curvature 1\,m and diameter $\sim$35\,cm is 
required. Figure~\ref{fig:CHEC} shows the camera structure and one of the 
photodetector modules. 
In addition to the elements listed CHEC will contain a calibration 
system based on LED flashers, reflecting light from the secondary mirror back 
on to the focal plane.

\subsection{Simulations}\label{sec:sims}
Simulations of CHEC performance have been made using CORSIKA output, assuming
perfect optics, and a custom electronics model. The following specifications 
are needed for CHEC-M to meet the CTA requirements for triggering: \newline 
{\bf Pulse Shape} (after pre-amplifier/at TARGET input): 10-90\%
rise-time 3.5-6.0 ns, and FWHM 5.5-10.5 ns. 
{\bf Trigger logic:} analogue sum of 4 pixels, discriminated and sent
to the camera trigger where a neighbour requirement and a minimum
multiplicity of 2 is applied within a coincidence window. 
{\bf Camera Trigger Coincidence Window:} 6-10\,ns. 
{\bf Time jitter on digital inputs to camera trigger:} $<2$\,ns for a
10\,ns coincidence window, $<1$\,ns for a 6\,ns window. 
{\bf Electronic Noise:} $<0.5$\,mV rms per pixel on the trigger path.
{\bf Pixel to pixel gain variations:} $<25$\% rms (4-pixel sum). 

\subsection{Photosensors}\label{sec:MAPM}

The CHEC-M detector plane will contain 32 H10966B MAPMs with Super-Bialkali 
photocathode, rotated and tiled to approximate the 1\,m radius of curvature 
optical focal plane. Each MAPM contain 64 pixels of size 
$\approx 6 \times 6$\,mm$^2$. This corresponds to an average angular size of 
$0.19^{\circ}$ when installed on ASTRI, and $0.17^{\circ}$ installed on GATE 
with the PSF ($\theta_{80}$) smaller than 6\,mm over the full CHEC field of 
view on both.
\begin{itemize}
\item {\bf Dynamic Range -} By 1000\,pe the response is 20\% non-linear. As 
  CHEC will provide full waveform data, useful information can still be 
  extracted from saturated channels via pulse fitting (to within 5\% at 
  1000\,pe).
\item {\bf Angular Response -} The SST-2M optics result in off-axis
  angles of up to 70$^\circ$ onto the focal plane. The angular response of a
  MAPM channel means at $70^\circ \sim 30$\% of the light is lost, but the
  incidence-angle averaged photon detection efficiency will be comfortably above
  that required over the whole camera field of view. 
\item {\bf Uniformity -} As there is a single HV supply for the 64 MAPM 
  pixels photon detection efficiency variations can not be removed and thus 
  affect the achievable dynamic range. 
  The effect on charge resolution is not expected to be 
  significant, provided the gain of each channel is measured to
  within $\sim$5\%. Such variations also affect the trigger threshold
  of the camera. One solution may be to include a variable input amplifier
  stage in the TARGET ASIC.
\item {\bf Ageing -} The gain of the MAPM pixels is expected to decline
  over time in relation to the integrated anode current in the device. 
  Tests show that for expected operating voltage and NSB level this effect
  is at an acceptable $<20$\% level over a decade of operation.
\end{itemize}
It is foreseen the MAPMs will be operated at a gain of 8$\times$10$^4$. 
The choice of SiPMs for CHEC-S will be made in Autumn 2013.

 \begin{figure*}[!t]
  \centering
  \includegraphics[width=\textwidth]{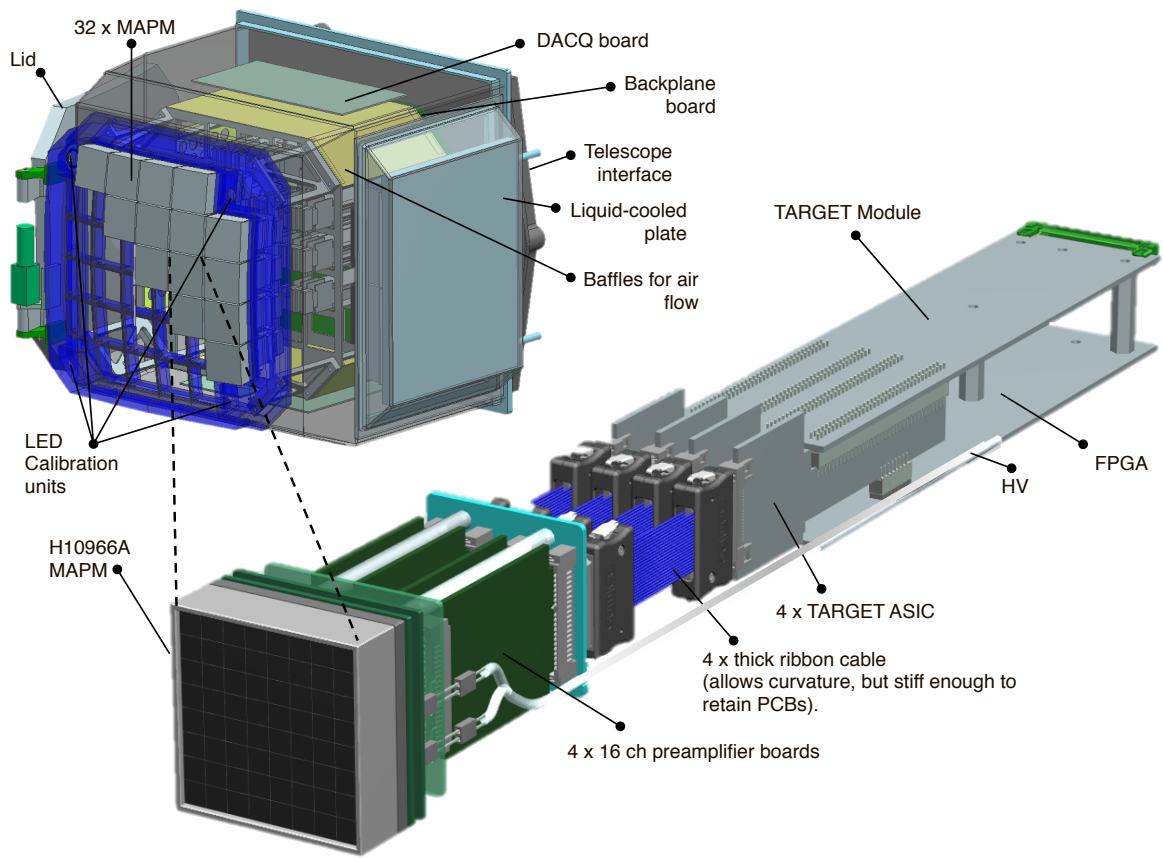}
  \caption{
           The CHEC camera mechanical structure and a TARGET Module.
          }
  \label{fig:CHEC}
 \end{figure*}

\subsection{Signal Amplification and Shaping}\label{sec:preamp}

The MAPMs must be operated at relatively low gain due to the high illumination 
levels from stars and the NSB. Preamplifiers are required in close proximity 
to the photosensors to minimise electronic noise and to shape the output 
signals for digitisation. 

CHEC-S will require a different preamplifier circuit to shape and
amplify SiPM signals to the required range. This development will
begin in full once the SiPM for CHEC-S has been selected, but still allow the
same interface to the digitisation electronics as CHEC-M.

\subsection{Digitisation}\label{sec:TARGET}

Digitisation of the amplified and shaped analogue signals will be
performed by the TARGET Modules developed at SLAC.
Each TARGET Module will supply a single MAPM with high voltage,
digitise the signals from all 64 channels, and provide these digitised
signals together with trigger information to the Backplane. These
modules are based on the TARGET ASIC \cite{bib:TARGET}. 
The shaped signal from the pre-amps will be digitised at 500-1000\,MSa/s over 
$\sim$100\,ns. The digitisation rate over this time results in an acceptable dead 
time for an expected camera trigger rate of $\sim$200\,Hz.
Above saturation the pulse area can be recovered by fitting the digitised 
waveform, to an accuracy of $\sim$5\% at 1000\,pe. 

\subsection{Trigger}\label{sec:trigger}

The TARGET ASIC provides the first level of triggering for the
camera. The trigger consists of the analogue sum of 4 neighbouring
pixels, which is then discriminated. A camera trigger requires any 2 
neighbouring trigger patches be present within a programmable coincidence time. 
External inputs to the FPGA from the DACQ board will also allow the camera to 
be artificially triggered. 
Due to the relatively low event rate per telescope, no
inter-telescope hardware array trigger is envisaged. When a telescope
triggers, all data is read out to the central location over
ethernet. Triggers are compared at the central location for several
telescopes to decide whether to write the data to disk. To make this
possible an accurate array-time distribution / event tagging scheme is
required. Within the CHEC prototype we plan to include 
an internal clock distribution and event-time-stamping system via a 
High Speed Deterministic Time Data Link (HSDTDL). 
The clock/array interface will be implemented on a daughter board and
connected directly to the Backplane. This provides flexibility for
changing the interface to match the final solution and/or try other options;  
for example, we are also considering the inclusion of a White Rabbit interface 
board \cite{bib:WhiteRabbit}.

\subsection{DACQ and Control}\label{sec:DACQ}

The TARGET Module serialises event data for output to the Backplane via a high 
speed data link (HSDL) comprised of differential RX and TX pairs. Serialisation 
and readout of 64 channels in this way minimises the connections to modules and 
when combined with the conversion time translates to a dead time well below the 
required 5\%. Both event data and control commands will be sent via this HSDL. 
The 32 RX and TX HSDLs from the TARGET Modules are routed to the DACQ
board via the Backplane. 
In the current model, the interface from the DACQ board to the `world'
would be via Gigabit Ethernet, 
providing sufficient resource for all data and control signals.

\subsection{Calibration}

 \begin{figure}[t]
  \centering
  \includegraphics[width=0.35\textwidth]{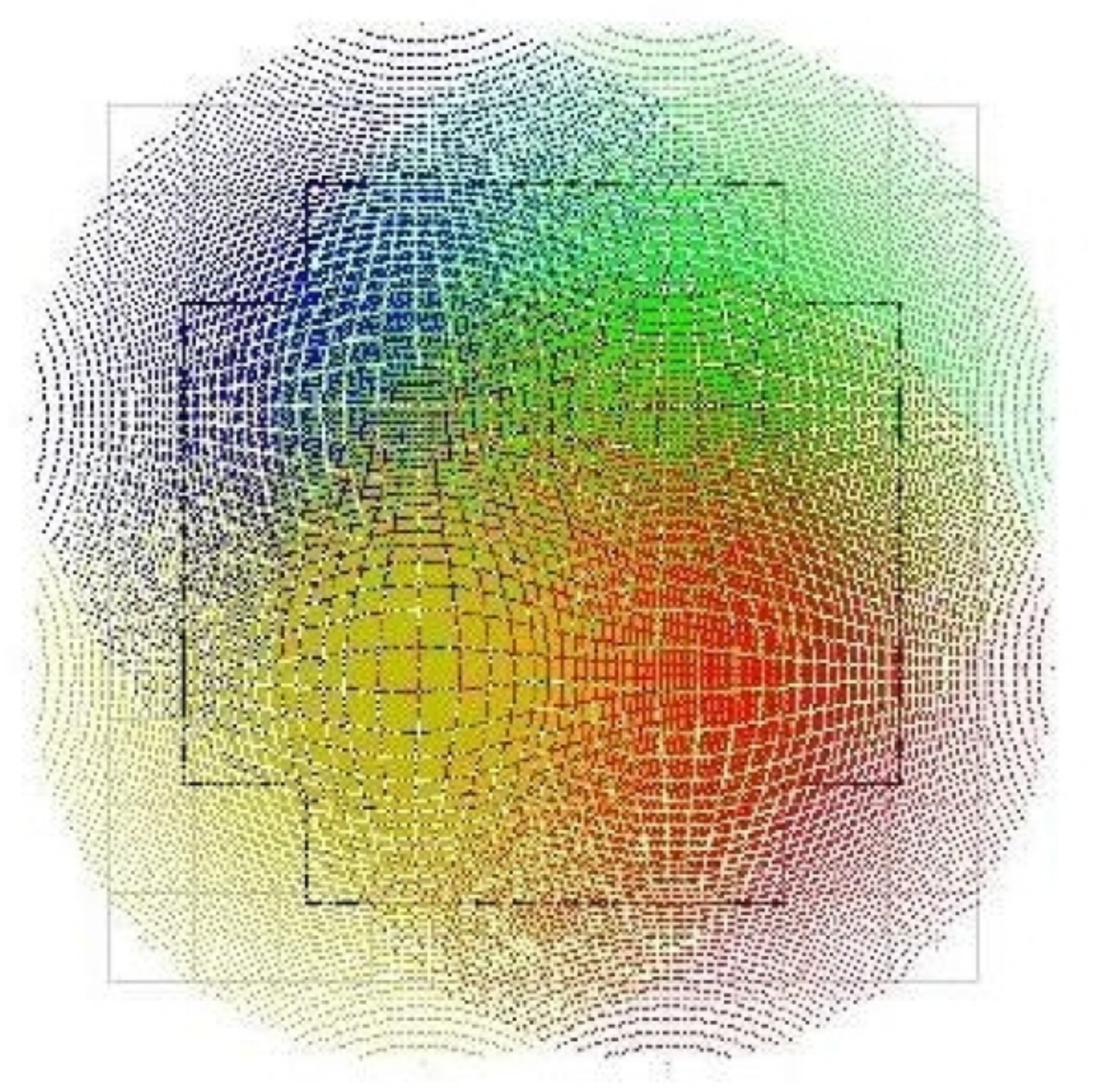}
  \caption{The calibration scheme, reflected light from each corner unit will 
           illuminate the entire focal plane.}
  \label{fig:flash}
 \end{figure}

CHEC includes an internal multiple LED calibration system with light pulse of
3-4\,ns (FWHM) at 400nm to flat-field the camera and
cover a large dynamic range of illuminations, from 0.1\,pe for
absolute single-pe calibration measurements, up to 1000\,pe to
characterise the camera up to and at saturation. The flasher units will be 
placed in the corners of the focal plane with diffusers to illuminate all pixels 
via reflection from the secondary mirror in a predictable way (see 
figure~\ref{fig:flash}. The flasher system has been integrated into the camera 
mechanical design.

At the beginning of each evening, the camera will be illuminated by the 
internal LED flashers at the single pe level whilst the telescope is at the 
park position. During normal observations calibration will be performed 
continuously for flat-field and dynamic range/linearity monitoring, accounting 
for any ambient effects.

\subsection{Mechanical structure and cooling}\label{sec:box}

A schematic of the camera structure is shown in Figure~\ref{fig:CHEC}. 
The mechanics for CHEC consist of: a focal-plane positioning plate,
internal `rack' mechanics, an adjustable interface plate for mounting
to the telescope, the external shell and lid to protect the focal plane 
when not observing. 

The focal-plane positioning plate located at the front of the camera
is responsible for the accurate positioning of the sensors. The TARGET
Modules with preamplifier modules attached, are slotted through this
plate and into the rack with the preamplifier module PCB flush to the surface 
and secured using screws. The MAPMs are then attached. The retention of the 
MAPMs will be provided by connectors reinforced with removable glue. A 
removable sealant will be used between each MAPM once they are attached to 
increase the retention and provide a seal against the elements.

For CHEC-S an `MAPM-like' module will be constructed, to minimise
changes to the mechanics. The pitch of the camera can be easily
changed to accommodate different physical sized sensor blocks of 64
pixels due to the flexible ribbon cables used to remove the radius of
curvature between the preamplifier and TARGET Modules.

Total power dissipation within the CHEC camera is expected to be $\leq 400$\,W. 
The resulting thermal control system consists of 4 fans coupled to a 
large water cooled heat sink. The fans, together with a system of 
baffles provide a recirculating airflow within the sealed camera 
enclosure.

To operate, CHEC requires power (12\,V) and a chilled water
supply. Both of these will be house in a `cabinet' located on the
telescope structure. A single multi-core fibre will carry data, control 
and clock signals. The software process controlling the camera is envisaged 
to run via a rack in the array control building.

\subsection{Operational Concept}

Due to the large number of SSTs and the anticipated long lifetime of CTA the 
reliability and maintainability requirements for CHEC will need to be superior 
to that of current IACT cameras, for which we are pursuing the following design 
routes:
\begin{itemize}
\item {\bf Removable Camera} as maintenance concept. A light camera designed to 
  be easily removable from the telescopes. The use of spare cameras and the 
  ability to inspect all elements of a camera for maintenance in an electronics
  workshop environment will allow repair of multiple issues
  simultaneously (and additional preventative maintenance) once a
  threshold in the number of inoperable pixels is reached.
\item {\bf Sealed system}. The cooling solution for CHEC does not
  involve circulation of air drawn from the external environment thus preventing
  the ingress of dust into the system and makes it much easier to 
  prevent the entrance of water/moisture into the system.
\item {\bf Extensive testing} at the prototyping stage. The test plan
  for CHEC is being designed to ensure that flaws in the design become
  apparent pre-mass-production. The accelerated ageing of electronics and 
  vibration and environmental testing of the camera as a whole (including 
  water-spray, salt-fog, hailstone impact tests, and thermal cycling) will 
  identify weaknesses in the design.
\end{itemize}

\section{Summary}

In summary, two CHEC prototypes will be constructed using almost identical 
electronics and different photosensors. CHEC-M will be installed on the ASTRI 
prototype structure in late 2014. CHEC-S will be ready $\sim$6\,months later. 
CHEC will include an internal LED flasher system for calibration of the
camera. The development of CHEC includes simulations of the trigger and readout 
performance to establish the specifications for hardware such that the CTA
requirements are met.

\vspace*{0.5cm}
\footnotesize{{\bf Acknowledgment:}
{
We gratefully acknowledge support from the agencies and organizations listed in 
this page: http://www.cta-observatory.org/?q=node/22.
We gratefully acknowledge financial support from the following agencies
and organisations: 
Science and Technology Facilities Council (STFC); University of Leicester; 
Netherlands Research School for Astronomy (NOVA), 
}
}

\end{document}